\documentclass{eas}
\usepackage{graphicx}
\usepackage{amssymb}
\begin{document}

\title{From Interstellar Clouds to Stars} 
\author{Jonathan C. Tan}\address{Depts. of Astronomy \& Physics, University of Florida, Gainesville, Florida 32611, USA}
\begin{abstract}
I review (1) Physics of Star Formation \& Open Questions; (2)
Structure \& Dynamics of Star-Forming Clouds \& Young Clusters; (3)
Star Formation Rates: Observations \& Theoretical Implications.
\end{abstract}
\maketitle
\section{The Physics of Star Formation and Open Questions}

Star formation is a complex, nonlinear phenomenon involving a host of
physical processes. However, in essence it is a competition between
self-gravity of gas clouds, along with any processes that promote
gravitational instability, and forces that resist collapse, including:
various pressures (e.g., thermal, turbulent, magnetic, cosmic ray);
direct stellar feedback (protostellar outflows, winds, ionization
[leading to enhanced thermal pressure], radiation pressure) and
rotation and/or shear. The evolution of these pressures and support
mechanisms needs to be followed by considering heating and cooling,
generation and decay of turbulence, and generation and diffusion of
$B$-fields. This requires following the chemical evolution of gas and
dust, both for heating/cooling and for the trace ionization fraction that
is important for coupling $B$-fields to the mostly neutral, molecular
gas.

We use {\it clumps} to describe self-gravitating gas structures that
fragment into star clusters and {\it cores} as the structures that
collapse via a central, rotationally-supported disk to form single
stars or small-$N$ multiples. Fragmentation of a clump into a
population of cores (and subsequent lack of fragmentation in core
envelopes) will depend on the evolution of local pressure support
contributions.

Once the clump or core contains stars, then, in addition to the
feedback these stars return to the gas, one also needs to follow the
evolution of their orbits and multiplicity properties in the
protocluster environment. Continued, competitive Bondi-Hoyle accretion
of gas to stars has been argued to be important in shaping the upper
end of the initial mass function (IMF) (Bonnell et al. 2001; Bate
2012). However, as discussed by Tan et al. (2014, hereafter T14),
simulations of competitive accretion have typically not included
$B$-fields or well-resolved feedback, especially protostellar
outflows, and so may have overestimated its importance.

This highlights two major challenges for star formation
simulations. First, one needs to be sure to include all the important
physical processes, especially $B$-fields (including non-ideal MHD
processes, since young stars have magnetic flux-to-mass ratios that
are several orders of magnitude smaller than molecular clouds) and
stellar feedback. Second, one needs to resolve a vast ranges of
scales, ideally down to the stellar surface. Unfortunately, this is
not yet practical for realistic systems that contain even moderate
numbers of stars, especially because of the short timesteps that are
associated with the smallest scales. Thus, subgrid models are always
needed, often implemented via sink particles, and a major concern is
whether the results of simulations, such as the IMF, binary properties
and star formation rates (SFRs), depend on how physics is implemented
at the subgrid level. A third challenge is, unlike the cosmological
Pop III case, numerical simulations of ``local'' star formation must
make uncertain choices for initial conditions.

Given these theoretical difficulties, and with observations of basic
quantities of star-forming regions (see \S\ref{S:Obs}), such as mass
and $B$-field strengths of interstellar gas clouds and the ages and
thus SFRs of young stellar systems, also suffering from large
uncertainties, there are still a number of major open questions about
the star formation process. On the scales of galactic disks and
kpc-scale subregions, empirical star formation relations have been
found between SFR per unit area, $\Sigma_{\rm SFR}$, and total gas
content per unit area, $\Sigma_g$, and orbital timescale (Kennicutt \&
Evans 2012), but, as discussed below in \S\ref{S:SFRs}, there is no
consensus on what processes regulate the SFR to follow these
relations. On scales of giant molecular clouds (GMCs), it is unclear
what causes a certain region to form stars, i.e., does it typically do
so because of an external trigger (e.g., converging flows, cloud
collisions or stellar feedback) or via spontaneous gravitationally
instability (e.g., as a cloud evolves and loses its earlier level of
internal pressure support)?  The answer to this question relates
directly to the specification of initial conditions for numerical
simulations of star formation, e.g., how close is the initial clump or
core to virial and pressure equilibrium?  How do stars accrete their
mass---is it mostly from a natal pre-stellar core or by competitive
Bondi-Hoyle accretion from the clump?  What sets the IMF and binary
properties and do these vary with environment?  What are the main
processes that regulate SFRs in GMCs, clumps \& cores: turbulence,
magnetic fields or feedback? Are timescales to form a star cluster
from a gas clump short (Elmegreen 2000; Hartmann \& Burkert 2007) or
long (Tan et al. 2006; Nakamura \& Li 2007) compared to the local
free-fall time?

\section{Structure and Dynamics of Star-Forming Clouds and Young Clusters}\label{S:Obs}

\begin{figure}[htb!]
\vspace{-0.3in}
\includegraphics[width=12cm]{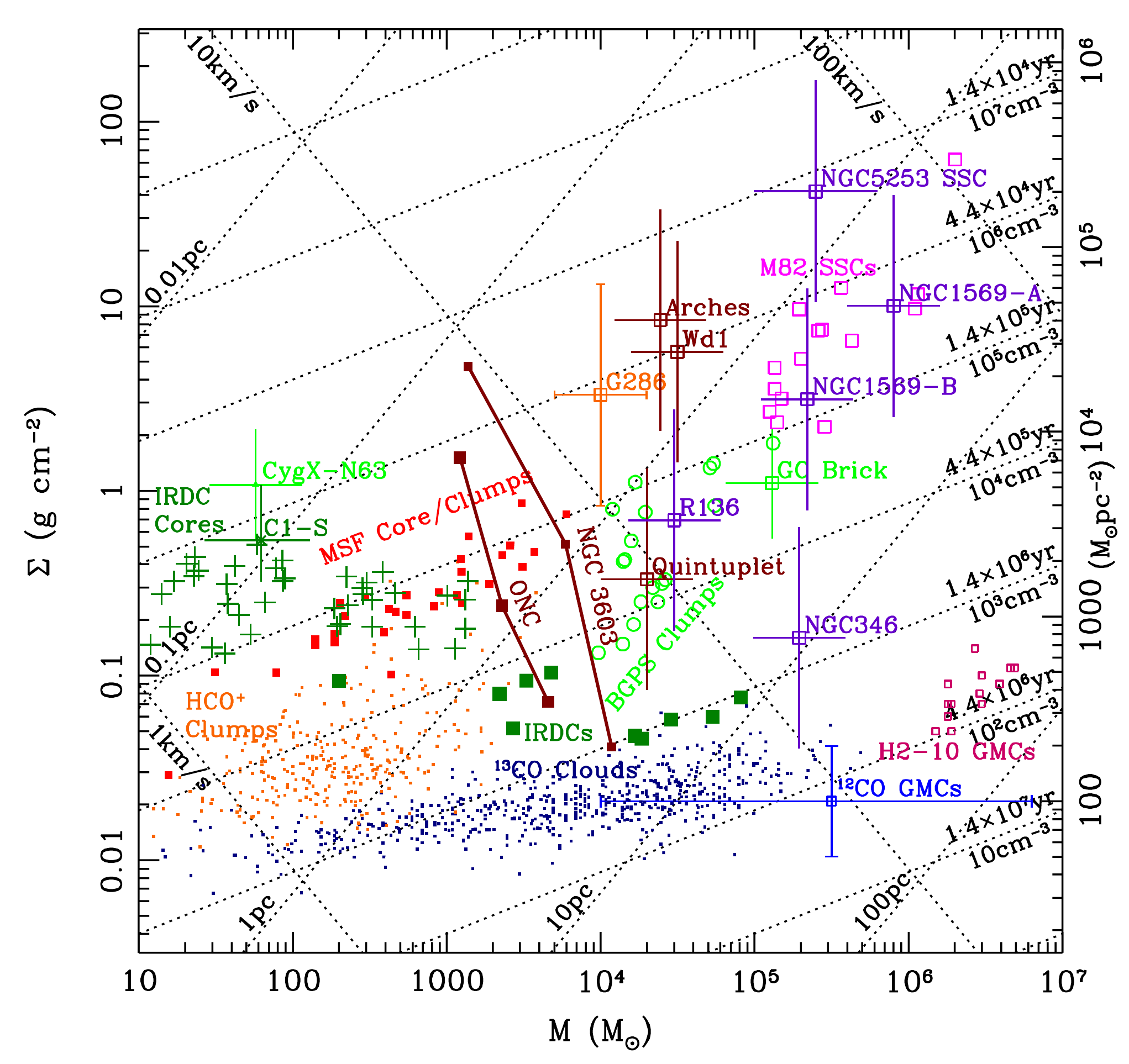}
\vspace{-0.25in}
\caption{
Physical properties of star-forming clouds and young clusters in the
Milky Way and nearby galaxies (from Tan et al. 2014). Mass surface
density, $\Sigma\equiv M/(\pi R^2)$, is plotted versus mass,
$M$. Dotted lines of constant radius, $R$, H number density, $n_{\rm
  H}$ (or free-fall time, $t_{\rm ff}= (3\pi/[32G\rho])^{1/2}$), and
escape speed, $v_{\rm esc} = (10/\alpha_{\rm vir})^{1/2}\sigma$, are
shown.
\label{fig:result}}
\end{figure}

The physical properties of mass ($M$), radius ($R$), mass surface
density ($\Sigma\equiv M/(\pi R^2)$), density ($\rho$) or equivalently
number density of H nuclei ($n_{\rm H}$), free-fall time ($t_{\rm
  ff}\equiv(3\pi/[32G\rho])^{1/2}$) and surface escape speed ($v_{\rm
  esc} = (10/\alpha_{\rm vir})^{1/2}\sigma$, where $\alpha_{\rm
  vir}\equiv5\sigma^2R/(GM)$ is the virial parameter and $\sigma$ is
the 1D velocity dispersion) of star-forming clouds and young star
clusters are shown in Fig. 1. The range of scales shown extends from
those of GMCs, i.e., up to $\sim 10^6\:M_\odot$ and $\sim100$~pc, down
to those of individual relatively massive cores $\sim 10\:M_\odot$ and
$\lesssim 0.1$~pc. 

A large fraction of stars in the Galaxy and other similar disk
galaxies are formed in star clusters with sizes of a few parsecs and a
range of $\Sigma$ from $\sim 0.1$ to $\sim 1\:{\rm g\:cm^{-2}}$ and
masses from $\sim 10^3$ to $\sim 10^5\:M_\odot$, with initial cluster
mass functions (ICMFs) that can be fit by a power law $dN/dM\propto
M^{-2}$ up to $\sim 10^5\:M_\odot$ (Fall \& Chandar 2012). Individual
galaxies with higher SFRs, such as the starbursting Antennae, or
samples of galaxies (e.g., Dowell et al. 2008) exhibit a similar ICMF
that extends up to $\gtrsim 10^6\:M_\odot$. Although the
majority of stars are born in protoclusters that have relatively high
overall star formation efficiencies, $\epsilon$, and are initially
gravitationally bound, most cluster members eventually disperse to
the field, due to a combination of cluster mass loss, envelope
expansion and tidal stripping.

The precursors to star clusters are dense clumps that are mostly
within GMCs and revealed by their emission from high critical density
species (e.g., HCO$^+$), from their extinction as Infrared Dark Clouds
(IRDCs), or by their mm/sub-mm dust continuum emission (e.g., Massive
Star-Forming (MSF) clumps and the Bolocam Galactic Plane Survey (BGPS)
clumps) (see Fig. 1). The central densities of some stellar clusters,
such as the Orion Nebula Cluster (ONC) and NGC 3603, extend to higher
values than are typically seen in gas clouds of equivalent mass, which
may indicate central concentration due to dynamical evolution of the
protocluster during or just after star cluster formation, probably
driven by mass segregation.

One important question is the extent to which GMCs and clumps are
gravitationally bound, i.e., with virial parameters $\alpha_{\rm
  vir}\lesssim 2$ and close to virial equilibrium, i.e., with
$\alpha_{\rm vir}\simeq 1$. Hernandez \& Tan (2015, hereafter HT15)
considered the ten IRDCs shown by solid green squares in Fig. 1, as
well as their surrounding GMCs as traced by $^{13}$CO(1-0). For clouds
defined as connected structures in position-position-velocity space
and with masses and velocity dispersions estimated from
$^{13}$CO(1-0), the mean and median values for GMCs are $\alpha_{\rm
  vir}=1.1$ and 1.0, with a standard deviation of 0.5. For the IRDCs,
the mean and median values are both 1.9, with a standard deviation of
1.1. This is evidence that GMCs are gravitationally bound and
virialized, consistent with the results of Roman-Duval et al. (2010,
see also Tan et al. 2013). It is tentative evidence for IRDCs having
more disturbed kinematics. However, one must be careful with
systematic effects, such as the possibility that CO freeze-out onto
dust grains is a greater in IRDCs (Hernandez et al. 2011), leading to
their $^{13}$CO-derived masses being somewhat underestimated.  As
pointed out by Heyer et al. (2009), studying the correlation of
$\sigma R^{-1/2}$ versus $\Sigma$, which should be linear if virial
equilibrium applies, is a more accurate way of assessing the virial
equilibrium of cloud populations, since it reduces the effects of
common systematic uncertainties associated with, e.g., mass
measurements that require assuming a particular abundance of a
molecular cloud tracer, such as $^{13}$CO. Such a correlation is seen
in the GMC sample studied by HT15, with a slight steepening on
extending into the higher $\Sigma$ regime of IRDCs.

Another way to assess the dynamics of clumps is to look directly for
infall motions and mass infall rates. As reviewed by T14, infall
times, $t_{\rm infall}\equiv M/\dot{M}_{\rm infall}$, relative to the
local free-fall time, $t_{\rm ff}$, have been measured in several
clumps, including IRDCs, with typical observed ratios of $\sim 10$,
indicating relatively slow, quasi equilibrium contraction. In
agreement with this, recently Wyrowski et al. (2015) have reported
infall velocities to a sample of nine massive molecular clumps,
finding infall speeds that are typically about 10\% of free-fall. Such
slow rates of collapse suggest either a dynamically important role for
$B$-fields, consistent with observations of two IRDCs by Pillai et
al. (2015, see also analysis of Falceta-Gon\c{c}alves et al. 2008 and H-B. Li et
al. 2014), and/or stabilization of collapse by protostellar outflow
feedback, as modeled by Nakamura \& Li (2007, 2014).

The timescales and dynamics of molecular clouds can also be assessed
by modeling their chemical evolution. From observations of ortho- and
para-$\rm H_2D^+$ that constrain the ortho-to-para ratio (OPR) of $\rm
H_2$, Br\"uncken et al. (2014) estimated a chemical age of $>1$~Myr,
i.e., $>10 t_{\rm ff}$, in the protostellar core IRAS 16293-2422 A/B,
which has mean density $n_{\rm H}\simeq 2.0\times 10^5\:{\rm
  cm^{-3}}$, i.e., $t_{\rm ff}\simeq 1.0\times 10^5$~yr. The OPR of
$\rm H_2$ also controls deuteration chemistry, e.g., of the abundance
ratio $D_{\rm frac}^{\rm N_2H^+}\equiv$[$\rm N_2D^+$]/[$\rm N_2H^+$]
(Pagani et al. 2013; Kong et al. 2015a). This ratio rises by several
orders of magnitude above the cosmic [D]/[H] ratio in cold, dense gas,
where OPR of $\rm H_2$ drops to very low values and CO molecules are
mostly frozen out onto dust grains. The study of Kong et al. (2015b)
uses observations of $D_{\rm frac}^{\rm N_2H^+}$ in two massive
starless or early-stage cores to estimate chemical ages that imply
contraction is most likely proceeding at rates smaller than $\sim1/3$
of free-fall.

Study of the internal structures of molecular clouds constrains the
processes that regulate their dynamics. A simple metric is the
probability distribution function (PDF) of $\Sigma$ within a defined
area. Kainulainen et al. (2009) found non-star-forming clouds have
log-normal $\Sigma$-PDFs, while star-forming clouds, such as Taurus,
have high-$\Sigma$ power-law tails. Log-normal $\Sigma$-PDFs are a
feature of supersonic turbulence, while high-$\Sigma$ power law tails
are seen to emerge in such simulations once self-gravity is turned on,
with this gas component interpreted as being in free-fall collapse
(Kritsuk et al. 2011; Collins et al. 2011). However, on the
observational side, the difficulty of determining the peak of
$\Sigma$-PDFs has been highlighted by Schneider et al. (2015a) and
Lombardi et al. (2015), with the latter concluding that, given the
uncertainties, the data of all the nearby clouds they considered are
consistent with pure power-law tails. On the other hand, from combined
NIR + MIR extinction mapping of an IRDC and surrounding GMC, Butler et
al. (2014) claimed to identify the peak of the $\Sigma$-PDF near
0.03~$\rm g\:cm^{-3}$ ($A_V\simeq 7$~mag) and found that the overall
shape was well-fit by a single log-normal, with very little scope for
the presence of high-$\Sigma$ power-law tail, even though the cloud
has $\alpha_{\rm vir}\simeq 1$ (however, see Schneider et
al. 2015b). On the theoretical side, simulations of self-gravitating,
strongly-magnetized (trans-Alfv\'enic), turbulent clouds with
realistic (i.e., non-periodic) boundary conditions are needed for
comparison with observed $\Sigma$-PDFs. In addition, the $\Sigma$-PDF
is just one simple metric and higher order spatial and/or kinematic
statistics are required to better characterize MHD turbulence
of molecular clouds (Burkhart et al. 2014; 2015).

Filamentary structures are common in star-forming clouds. On the
largest scales, at least ten examples of very long $\sim100$~pc-scale
filaments have been detected in the Galactic plane (e.g., Jackson et
al. 2010; Ragan et al. 2014; Wang et al. 2015). Their relatively
ordered kinematics are a challenge to models that involve rapid,
free-fall collapse (Butler et al. 2015), which may again indicate that
$B$-field support is important. On smaller scales, molecular clouds
and clumps often appear to be composed of a network of filaments and
there are claims, from studies of nearby, relatively low-$\Sigma$
clouds, that the filaments exhibit a characteristic width of
$\sim0.1$~pc (see review by Andr\'e et al. 2014). Hennebelle \&
Andr\'e (2013) have presented a model of self-gravitating filaments in
which this scale is explained as being due to a balance between
accretion-driven turbulence and dissipation by ion-neutral friction.
More generally, sheets and filaments are expected to occur in all
realistic models involving gravitational collapse, including those in
which the clouds are permeated by trans-Alfv\'enic turbulence for
which both $B$-fields and turbulent motions make important
contributions to internal pressure support.

Both pre-stellar and protostellar cores are seen in star-forming
clouds, typically spaced along filaments, with identification via dust
continuum emission, dust extinction or molecular emission lines. The
core mass function (CMF) has been measured and found to be similar in
shape to the stellar IMF (e.g., K\"onyves et al. 2015), but translated
to higher masses by a factor of about 2.5, i.e., implying a star
formation efficiency of about 40\%. Such efficiencies are naturally
explained as being due to feedback from protostellar outflows, for
both low-mass (Matzner \& McKee 2000) and high-mass cores (Zhang et
al. 2014). 

The origin of the CMF remains uncertain. Models of
turbulent fragmentation have been proposed. For example, Padoan \&
Nordlund (2002) build a CMF by equating the size of cores to the
thickness of post-shock compressed gas in a cloud permeated by
super-Alfv\'enic turbulence (see also Hennebelle \& Chabrier
2008). However, it is not clear that observed cores form in such a
dynamic and rapid manner. Alternatively, Kunz \& Mouschovias (2009)
have proposed a dominant role for $B$-fields, with the CMF being set
by ambipolar diffusion.

The velocity dispersions of core populations have been measured in
star-forming regions: e.g., the $\rm N_2H^+$(1-0)-defined sample of
Kirk et al. (2007) in Perseus. It was noted that the velocity
dispersions appear subvirial, which could either be a true reflection
of an unstable dynamical state (although dense cores are expected to
exhibit somewhat subvirial motions even in virialized clouds, Offner
et al. 2008) or indicate that large-scale $B$-fields are providing
significant support in the clump.

Observations of the stellar kinematics of embedded clusters from NIR
radial velocities find a dynamical state that is near virial
equilibrium in the case of NGC 1333 (Foster et al. 2015) and moderately
super-virial for IC 348 (Cottaar et al. 2015). These
results are consistent with theoretical expectations, given that the
latter cluster has a smaller gas mass fraction, perhaps due to
dispersal by feedback.

\section{Star Formation Rates - Observations and Theoretical Implications}\label{S:SFRs}

It is very challenging to measure SFRs. On galactic and
kiloparsec-scales this is normally achieved by measuring diagnostics
associated with short-lived (i.e., $\lesssim 30$~Myr) high-mass stars,
such as recombination lines, mainly H$\alpha$, from ionized gas and
MIR to FIR luminosities (see Kennicutt \& Evans 2012). Empirically, a
correlation $\Sigma_{\rm SFR} = (6.3\pm1.8)\times10^{-3}
(\Sigma_g/10\:M_\odot\:{\rm pc}^{-2})^{1.4\pm0.15}\:M_\odot\:{\rm
  yr}^{-1}\:{\rm kpc}^{-2}$ is seen in disk galaxies and circumnuclear
starburst disks, i.e., with $10 < \Sigma_g/(M_\odot\:{\rm
  pc}^2)\lesssim 10^5$ (Kennicutt 1998). Considering only the mass
surface density of molecular gas, $\Sigma_{\rm H2}$, a linear relation
$\Sigma_{\rm SFR} = (5.3\pm0.3)\times10^{-3} (\Sigma_{\rm
  H2}/10\:M_\odot\:{\rm pc}^{-2})\:M_\odot\:{\rm yr}^{-1}\:{\rm
  kpc}^{-2}$ has been derived (Bigiel et al. 2008). Another relation
that is almost as good a fit to the data is $\Sigma_{\rm SFR}\simeq
6\times 10^{-3}\Sigma_g\Omega$, which is equivalent to conversion of
4\% of total gas mass into stars per local orbital time $t_{\rm
  orb}=2\pi/\Omega$ (Kennicutt 1998; Tan 2010; Suwwanajak et
al. 2014). Such a relation could in principle be explained by
triggering of star formation by spiral arm passage, but such
enhancements were not seen in the study of Foyle et al. (2010). One
alternative model is triggering by shear-driven GMC collisions in a
flat rotation curve thin disk (Tan 2000; Tasker \& Tan 2009), since in
this environment, for gravitationally bound GMCs, the collision time is
a fixed fraction, $\sim 1/5$, of an orbital time.

Once the total SFR is measured in a galaxy, such as the Milky Way
where it is estimated to be ${\rm SFR}_{\rm tot}\sim 3\:M_\odot\:{\rm
  yr^{-1}}$, then the SFR per free-fall time, $\epsilon_{\rm ff}$, on
the density scale of a given ISM tracer can also be assessed. For
example, the Galactic population of $^{12}$CO-defined GMCs has $M_{\rm
  tot}\sim 10^9\:M_\odot$ with typical densities $n_{\rm
  H}\sim10^2\:{\rm cm^{-2}}$ and free-fall times $\sim 4\times
10^6$~yr. Assuming most of the total SFR occurs in GMCs, then
$\epsilon_{\rm ff}= {\rm SFR_{\rm tot}} t_{\rm ff}/M_{\rm tot}
\rightarrow 0.01$ for GMCs (Zuckermann \& Evans 1974). Extending this
to the higher densities of IRDCs and other tracers, Krumholz \& Tan
(2007) concluded $\epsilon_{\rm ff}\sim 0.02$ for clouds of densities
up to $\sim 10^5\:{\rm cm^{-3}}$. Murray (2011) found $\sim10\times$
higher values, i.e, $\epsilon_{\rm ff}\sim 0.2$, in the GMCs
associated with the 13 most luminous Galactic free-free sources, which
suggests that SFR activity may be very stochastic.

Extending to the smaller scales of individual GMCs, counting of young
stellar objects (YSOs) can be attempted and a SFR derived by assuming
they have a typical age, e.g., $\sim 2\pm1$~Myr in the study of
Heiderman et al. (2010). Their results are consistent with a threshold
for star formation of $\Sigma_{\rm th}\simeq
130\:M_\odot\:{\rm pc}^{-2}$, i.e., $\simeq 0.03\:{\rm g\:cm^{-2}}$ or
$A_V\simeq 7$~mag, which agrees with the prediction of a model
of photoionization-regulated star formation in magnetically-supported
clouds (McKee 1989). Above the threshold, there is an approximately
linear relation of $\Sigma_{\rm SFR} \simeq 0.16 (\Sigma_{\rm
  H2}/10\:M_\odot\:{\rm pc}^{-2})^{1.1}\:M_\odot\:{\rm yr}^{-1}\:{\rm
  kpc}^{-2}$, i.e., with a normalization factor $\sim 30\times$
higher than the kpc-scale result of Bigiel et al. (2008). This implies
$\epsilon_{\rm ff}= 0.05 (M/10^5\:M_\odot)^{1/4}
(\Sigma/200\:M_\odot{\rm pc}^{-2})^{-0.65}$ for spherical clouds.

If more accurate individual stellar ages can be derived,
then age spreads can be searched for and SFRs derived more
directly. The best region for such studies is the ONC, from which an
estimate of $\epsilon_{\rm ff}\sim 0.04$ is derived (Da Rio et
al. 2014).

Ma et al. (2013) measured luminosity to mass ratios, $L/M$, of a
complete census of HCO$^+$ clumps (Fig. 1). The lower bound of this
distribution in the $(L/M)$ vs. $\Sigma$ plane constrains
$\epsilon_{\rm ff}\lesssim 0.2$, otherwise just the accretion
luminosities of the protostars, i.e., ignoring internal luminosities,
would exceed observed values.

Thus, even in regions that are actively forming stars, observed SFRs
in molecular clouds typically imply very inefficient star formation,
i.e., just a few percent conversion of gas into stars per local
free-fall time. Based on the results of numerical simulations,
supersonic turbulence in self-gravitating gas has been proposed to
explain these SFRs (Krumholz \& McKee 2005; Padoan \& Nordlund 2011;
Hennebelle \& Chabrier 2011; see review by Padoan et
al. 2014). Krumholz et al. (2009) have applied these models to
populations of GMCs in galactic disks to explain their empirical star
formation relations.

However, the turbulence regulated star formation models involve
creation of gravitationally unstable regions, i.e., cores, by rapid
compression behind shocks and it is not clear that the observations of
cores (\S\ref{S:Obs}) support a picture of such rapid accumulation of
their gas. Also, to achieve $\epsilon_{\rm ff}<0.05$ in simulations of
MHD turbulence, typically requires $\alpha_{\rm vir}>6$ (Padoan et
al. 2012), again inconsistent with observed clouds. The simulations
that underpin the models are mostly of super-Alfv\'enic turbulence
(i.e., relatively weak $B$-fields) driven on large-scales and have
periodic boundary conditions. If star cluster formation is a
relatively slow, quasi-equilibrium process that takes at least several
free-fall times to build up the stellar population (Tan et al. 2006),
then most stars will instead be forming in an environment in which
turbulence is maintained by small-scale driving by protostellar
outflows (Nakamura \& Li 2007).  Finally, the $\Sigma$
threshold for star formation is not naturally explained in these pure
turbulence models.

In summary, observations suggest stars form with low efficiency per
local free-fall time, i.e., $\epsilon_{\rm ff}$ of a few percent, in
clumps with $A_V\gtrsim 7$~mag, i.e., $\Sigma\gtrsim 0.03\:{\rm
  g\:cm^{-2}}$, but these regions often eventually build up overall
efficiencies of $\epsilon\gtrsim 0.2$ to form loosely bound
clusters. This star formation activity is likely regulated by a
combination of $B$-fields and turbulence. Clumps need to become
moderately magnetically supercritical, i.e., their large-scale
$B$-fields no longer strong enough to resist gravity, which likely
explains the $\Sigma$ threshold for active SFRs. However, $B$-fields
remain relatively strong so that turbulence is trans-Alfv\'enic. At
first turbulence is mostly driven from large scales, perhaps by GMC
collisions, which also promote compression and formation of
supercritical clumps and are a natural mechanism to connect the
kpc-scale galactic dynamics of shearing disks to the pc-scales of
protoclusters. As star formation proceeds, then the local turbulence
becomes dominated by that driven by protostellar outflows. Eventually
feedback from massive stars, especially ionization, disperses
remaining gas in the protocluster, though this becomes progressively
more difficult in higher-$\Sigma$, higher-mass clumps, which may then
form super star clusters with high efficiency.



\begin{thebibliography}{99}

\bibitem[]{}
Andr\'e, P., Di Francesco, J., Ward-Thompson, D. et al. 2014, {\it Protostars \& PlanetsVI}, 27

\bibitem[]{}
Bate, M. R. 2012, \textit{MNRAS}, 419, 3115

\bibitem[]{}
Bigiel, F., Leroy, A., Walter, F. et al. 2008, {\it AJ}, 136, 2846

\bibitem[Bonnell et al.(2001)]{2001MNRAS.323..785B} 
Bonnell, I.~A., Bate, M.~R., Clarke, C.~J., \& Pringle, J.~E.\ 2001, {\it MNRAS}, 323, 785

\bibitem[]{}
Br\"uncken, S., Sipil\"a, O., Chambers, E. T. et al. 2014, {\it Nature}, 516, 219

\bibitem[]{}
Burkhart, B., Collins, D. C., \& Lazarian, A. 2015, {\it ApJ}, 805, 118

\bibitem[]{}
Burkhart, B., Lazarian, A., Le\~ao, I. C. et al. 2014, {\it ApJ}, 790, 130

\bibitem[]{}
Butler M. J., Tan J. C., \& Kainulainen, J. 2014, {\it ApJL}, 782, L30

\bibitem[]{}
Butler M. J., Tan J. C., \& Van Loo, S. 2015, {\it ApJ}, 805, 1

\bibitem[]{}
Collins, D., Padoan, P., Norman, M. L., \& Xu, H. 2011, \textit{ApJ}, 731, 59

\bibitem[]{}
Cottaar, M., Covey, K. R., Foster, J. B. et al. 2015, \textit{ApJ}, 807, 1

\bibitem[]{}
Da Rio, N., Tan, J. C., \& Jaehnig, K. 2014, \textit{ApJ}, 795, 55

\bibitem[]{}
Dowell, J. D., Buckalew, B. A., \& Tan, J. C. 2008, {\it AJ}, 135, 823

\bibitem[]{}
Elmegreen, B. G. 2000, \textit{ApJ}, 530, 277

\bibitem[]{}
Falceta-Gon\c{c}alves, D., Lazarian, A. \& Kowal, G. 2008, {\it ApJ}, 679, 537

\bibitem[]{}
Fall, S. M., \& Chandar, R. 2012, \textit{ApJ}, 752, 96

\bibitem[]{}
Foster, J. B., Cottaar, M., Covey, K. R. et al. 2015, \textit{ApJ}, 799, 136

\bibitem[]{}
Foyle, K., Rix, H.-W., Walter, F., \& Leroy, A. K. 2010, {\it ApJ}, 725, 534

\bibitem[]{}
Hartmann, L., \& Burkert, A. 2007, \textit{ApJ}, 654, 988

\bibitem[]{}
Heiderman, A., Evans, N. J., Allen, L. E. et al. 2010, {\it ApJ}, 723, 1019

\bibitem[]{}
Hennebellw, P., \& Andr\'e, P. 2013, {\it A\&A}, 560, 68

\bibitem[]{}
Hennebelle, P. \& Chabrier, G. 2008, \textit{ApJ}, 684, 395

\bibitem[]{}
Hennebelle, P. \& Chabrier, G. 2011, \textit{ApJL}, 743, L29

\bibitem[]{}
Hernandez, A. K., \& Tan, J. C. 2015, \textit{ApJ}, 809, 154

\bibitem[]{}
Hernandez, A. K., Tan, J. C., Caselli et al. 2011, {\it ApJ}, 738, 11

\bibitem[]{}
Jackson, J. M., Finn, S. C., Chambers, E. T. et al. 2010, {\it ApJL}, 719, L185

\bibitem[]{}
Kainulainen, J., Beuther, H., Henning, T., \& Plume, R. 2009, {\it A\&A}, 508, L35

\bibitem[]{}
Kennicutt, R. C. 1998 {\it ApJ}, 498, 541

\bibitem[]{}
Kennicutt, R. C., \& Evans, N. J. 2012, {\it ARA\&A}, 50, 531

\bibitem[]{}
Kirk, H., Johnstone, D., \& Tafalla, M. 2007, {\it ApJ}, 668, 1042


\bibitem[]{}
Kong, S., Caselli, P., Tan, J. C. et al. 2015a, \textit{ApJ}, 804, 98

\bibitem[]{}
Kong, S., Tan, J. C., Caselli, P. et al. 2015b, \textit{ApJ}, submitted (arXiv:1509.08684)

\bibitem[]{}
K\"onyves, V., Andr\'e, P., Men'shchikov, A. et al. 2015, {\it A\&A}, 548, 91

\bibitem[]{}
Kritsuk, A. G., Norman, M. L., \& Wagner, R. 2011, {\it ApJ}, 727, L20

\bibitem[]{}
Krumholz, M. R. \& McKee, C. F. 2005, {\it ApJ}, 630, 250

\bibitem[]{}
Krumholz, M. R., McKee, C. F., \& Tumlinson, J. 2009, {\it ApJ}, 699, 850


\bibitem[]{}
Krumholz, M. R. \& Tan, J. C. 2007, {\it ApJ}, 654, 304

\bibitem[]{}
Kunz, M. W. \& Mouschovias, T. Ch. 2009, \textit{MNRAS}, 399, L94

\bibitem[]{}
Li, H.-B., Goodman, A., Sridharan, T. K. et al. 2014, {\it Protostars \& Planets VI}, 101

\bibitem[]{}
Lombardi, M., Alves, J., \& Lada, C. J. 2015, {\it A\&A}, 576, L1

\bibitem[]{}
Ma, B., Tan, J. C., \& Barnes, P. 2013, {\it ApJ}, 779, 79

\bibitem[]{}
Matzner, C. D. \& McKee, C. F. 2000, {\it ApJ}, 545, 364

\bibitem[]{}
McKee, C. F. 1989, {\it ApJ}, 345, 782

\bibitem[]{}
Murray, N. 2011, {\it ApJ}, 729, 133

\bibitem[]{}
Nakamura, F. \& Li, Z.-Y. 2007, \textit{ApJ}, 662, 395

\bibitem[]{}
Nakamura, F. \& Li, Z.-Y. 2014, \textit{ApJ}, 783, 115

\bibitem[]{}
Offner, S. S. R., Krumholz, M. R., Klein, R. I. \& McKee, C. F. 2008, {\it AJ}, 136, 404

\bibitem[]{}
Padoan, P., Federrath, C., Chabrier, G. et al. 2014, \textit{Protostars \& Planets VI}, p77

\bibitem[]{}
Padoan, P., Haugb{\o}lle, T., \& Nordlund, \AA. 2012, {\it ApJ}, 759, L27

\bibitem[]{}
Padoan, P. \& Nordlund, \AA. 2002, \textit{ApJ}, 576, 870

\bibitem[]{}
Padoan, P. \& Nordlund, \AA. 2011, \textit{ApJ}, 730, 40

\bibitem[]{}
Pagani, L., Lesaffre, P. Jorfi, M. et al. 2013, {\it A\&A}, 551, 38

\bibitem[Pillai et al.(2015)]{2015ApJ...799...74P} 
Pillai, T., Kauffmann, J., Tan, J.~C., et al. 2015, {\it ApJ}, 799, 74

\bibitem[]{}
Ragan, S. E., Henning, Th., Tackenberg, J., et al. 2014, {\it A\&A}, 568, A73

\bibitem[]{}
Roman-Duval, J., Jackson, J. M., Heyer, M. et al. 2010, {\it ApJ}, 723, 492

\bibitem[]{}
Schneider, N., Ossenkopf, V., Csengeri, T. et al. 2015a, {\it A\&A}, 575, 79

\bibitem[]{}
Schneider, N., Csengeri, T., Klessen, R. S. et al. 2015b, {\it A\&A}, 578, 29

\bibitem[]{} 
Suwwanajak, C., Tan, J. C., \& Leroy, A. K. 2014, {\it ApJ}, 787, 68

\bibitem[]{} 
Tan, J. C. 2000, \textit{ApJ}, 536, 173

\bibitem[]{} 
Tan, J. C. 2010, \textit{ApJ}, 710, 88

\bibitem[Tan et al.(2014)]{2014prpl.conf..149T} 
Tan, J.~C., Beltr{\'a}n, M.~T., Caselli, P., et al.\ 2014, {\it Protostars \& Planets VI}, 149

\bibitem[]{} 
Tan, J. C., Krumholz, M. R., \& McKee, C. F. 2006, \textit{ApJL}, 641, L121

\bibitem[]{} 
Tan, J. C., Shaske, S., \& Van Loo, S. 2013, {\it IAU Symp. 292}, ed. T. Wong \& J. Ott, 19

\bibitem[]{} 
Tasker, E. J., \& Tan 2009, {\it ApJ}, 700, 358

\bibitem[]{} 
Wang, K., Testi, L., Ginsburg, A. et al. 2015, {\it MNRAS}, 450, 4043

\bibitem[]{}
Wyrowski, F., G\"usten, R., Menten, K. M. et al. 2015, {\it A\&A}, in press (arXiv:1510.08374)

\bibitem[]{}
Zhang, Y., Tan, J. C., \& Hosokawa, T. 2014, \textit{ApJ}, 788, 166

\bibitem[]{}
Zuckerman, B., \& Evans, N. J. 1974, {\it ApJL}, 192, L194


\end{thebibliography}
\end{document}